\documentclass[conference]{IEEEtran}
\IEEEoverridecommandlockouts
\usepackage{cite}
\usepackage{amsmath,amssymb,amsfonts}
\usepackage{graphicx}
\usepackage{booktabs}
\usepackage{url}
\usepackage{textcomp}
\usepackage{xcolor}
\usepackage{orcidlink}
\IEEEoverridecommandlockouts\IEEEpubid{\makebox[\columnwidth]{979-8-3195-0662-7/26/\$31.00 $\copyright$2026 IEEE \hfill}\hspace{\columnsep}\makebox[\columnwidth]{ }}

\begin{document}

\title{Sentinel-Based Failover for QKD-Augmented IPsec Tunnels}

\author{\IEEEauthorblockN{1\textsuperscript{st} Juan Carlos Hernandez-Hernandez \orcidlink{0000-0003-1854-2070}}
\IEEEauthorblockA{
\textit{University of Luxembourg}\\
Luxembourg City, Luxembourg\\
juancarlos.hernandez@uni.lu}
\and
\IEEEauthorblockN{2\textsuperscript{nd} Francesco Vista \orcidlink{0000-0002-2728-0642}}
\IEEEauthorblockA{
\textit{University of Luxembourg}\\
Luxembourg City, Luxembourg\\
francesco.vista@uni.lu}
\and
\IEEEauthorblockN{3\textsuperscript{rd} Haftay Gebreslasie Abreha
\orcidlink{0009-0000-2172-2326}} 
\IEEEauthorblockA{
\textit{University of Luxembourg}\\
Luxembourg City, Luxembourg\\
haftay.abreha@uni.lu}
\and
\IEEEauthorblockN{4\textsuperscript{th} Intidhar Bedhief
\orcidlink{0000-0001-7921-2572}}
\IEEEauthorblockA{
\textit{University of Luxembourg}\\
Luxembourg City, Luxembourg\\
intidhar.bedhief@uni.lu}
\and
\IEEEauthorblockN{5\textsuperscript{th} Seid Koudia
\orcidlink{0000-0002-7533-6778}} 
\IEEEauthorblockA{
\textit{University of Luxembourg}\\
Luxembourg City, Luxembourg\\
seid.koudia@uni.lu}
\and
\IEEEauthorblockN{6\textsuperscript{th} Symeon Chatzinotas \orcidlink{0000-0001-5122-0001}}
\IEEEauthorblockA{
\textit{University of Luxembourg}\\
Luxembourg City, Luxembourg\\
symeon.chatzinotas@uni.lu}
}

\maketitle

\begin{abstract}
Quantum-safe IPsec through hybrid key establishment is practical, but creates a critical operational challenge: how to maintain tunnel availability when the QKD infrastructure becomes unavailable. In this paper, we present the design, implementation, and experimental evaluation of a quantum-safe key establishment mechanism for an IPsec tunnel that combines X25519, ML-KEM, and ETSI GS QKD 014 keys through the RFC 9370 multiple key exchange mechanism, and that degrades gracefully when the QKD key delivery fails. Our open-source StrongSwan plugin uses a sentinel-based coordination protocol, thereby permitting us to complete the handshake even if the QKD leg fails, instead of aborting, restoring the QKD share at the next rekey. On a testbed connected to a metropolitan QKD link over 33 km of deployed fiber, we evaluated five configurations, from a classical X25519 with RSA baseline to a hybrid one that adds ML-KEM-1024 and a QKD key. The full hybrid authentication costs 103 ms against 61 ms for the baseline, the QKD retrieval itself adds only about 7 ms. Failure injection experiments confirm that the tunnel survives a complete KME outage without any interruption of the protected traffic.
\end{abstract}


\section{Introduction}
IPsec constitutes one of the most widely deployed mechanisms to protect traffic at the network layer. The keys used to encrypt information are negotiated with the Internet Key Exchange protocol version 2 (IKEv2) defined in RFC 7296~\cite{rfc7296}, based on classical public key algorithms. These algorithms such as Elliptic Curve Diffie-Hellman and RSA, are exactly those that a cryptographically relevant quantum computer would break with Shor's algorithm. However, the threat is more urgent with the ``Harvest Now, Decrypt Later'' attack, an attacker can record encrypted traffic today and decrypt it once a quantum computer becomes available.

Two complementary defenses have emerged. Post-Quantum Cryptography (PQC) replaces the vulnerable algorithms with new ones, such as the standardized ML-KEM key encapsulation mechanism~\cite{fips203}. Quantum Key Distribution (QKD) enables symmetric keys whose secrecy rests on quantum physics. The IETF has extended IKEv2 so that both defenses can be used at the same time. We can highlight RFC 9242~\cite{rfc9242} which adds an intermediate exchange for large payloads, and RFC 9370~\cite{rfc9370} allowing several key exchanges (KEs) inside one negotiation, so classical, PQC, and QKD derived secrets can all feed a single session key. The resulting tunnel stays secure as long as at least one component remains unbroken~\cite{rfc9370}. This strategy is relevant in transition periods such as the one we are running today, where the migration from classical cryptography to quantum-safe implies maturity and adoption of the news schemes.  

However, the hybridization creates a new operational dependency. Enabling QKD is not as easy as doing local computation as PQC. It requires a live connection to a Key Management Entity (KME) that serves keys produced by the quantum link. KMEs can fail because of either key exhaustion or becoming unreachable for any reason. If the IKEv2 negotiation simply aborts in these cases, adding QKD would reduce the availability of the tunnel instead of improving its security. This trade off between defense in depth and availability is, in our view, the main practical obstacle for QKD adoption in IPsec, and it has received much less attention than raw performance such as quantum bit error or secret key rate.

\begin{table*}[!t]
\caption{Comparison of Related Work on Quantum-Safe IPsec}
\label{tab:related}
\centering
\begin{tabular}{lccccc}
\toprule
\textbf{Work} & \textbf{Year} & \textbf{PQC} & \textbf{QKD} & \textbf{IKEv2/IPsec} & \textbf{Graceful degradation} \\
\midrule
Herzinger \emph{et al.}~\cite{herzinger2023realworld} & 2021 & \checkmark & -- & \checkmark &  -- \\
Dervisevic and Mehic~\cite{dervisevic2021qkdipsec} & 2021 & -- & \checkmark & \checkmark & -- \\
Alia \emph{et al.}~\cite{alia2024deployedfiber} & 2024 & -- & \checkmark & \checkmark & -- \\
Twardokus \emph{et al.}~\cite{twardokus2025viability} & 2025 & \checkmark & -- & \checkmark & -- \\
Blanco-Romero \emph{et al.}~\cite{blanco2025hybrid} & 2025 & \checkmark & \checkmark & \checkmark & -- \\
Iliadis-Apostolidis \emph{et al.}~\cite{iliadis2025qrons} & 2025 & \checkmark & -- & \checkmark & -- \\
Vicente \emph{et al.}~\cite{vicente2026banking} & 2026 & \checkmark & \checkmark & \checkmark & -- \\
\textbf{This work} & 2026 & \checkmark & \checkmark & \checkmark & \checkmark \\
\bottomrule
\end{tabular}
\end{table*}

In this paper we answer the question of how can the tunnel keep working when the QKD key delivery fails, without manual intervention and without renegotiating from scratch? We make the following contributions:
\begin{itemize}
    \item We design and implement an open source strongSwan plugin\footnote{\url{https://github.com/jhhdez/qkdPlugin}} that integrates ETSI GS QKD 014 key retrieval as an additional RFC 9370 key exchange, in addition to X25519 and ML-KEM.
    \item We propose and validate a sentinel based graceful degradation protocol to keep the tunnel up and re-keying successfully in case of any KME failure.
    \item We evaluate five IKEv2 configurations, from classical X25519 with RSA to X25519 with ML-KEM-1024 and a QKD key, on a testbed attached to a metropolitan QKD link over 33 km of deployed fiber.
\end{itemize}

The rest of the paper is organized as follows. Section~\ref{sec:background} gives the necessary background. Section~\ref{sec:sota} reviews the state of the art. Section~\ref{sec:design} presents the system design and the degradation protocol. Section~\ref{sec:setup} describes the experimental setup, and Section~\ref{sec:results} the results. Section~\ref{sec:conclusion} concludes the paper.

\section{Background}
\label{sec:background}

\subsection{IKEv2 and the Rekeying Discipline}
IPsec protects IP traffic with the Encapsulating Security Payload (ESP) with IKEv2~\cite{rfc7296} negotiating the required Security Associations (SAs). A standard negotiation consists of an IKE\_SA\_INIT exchange, which performs the key exchange, followed by an IKE\_AUTH exchange, which authenticates the peers and creates the first Child SA. Long lived tunnels renew their keys periodically with CREATE\_CHILD\_SA exchanges, both for the IKE SA (IKE\_SA\_REKEY) and for the Child SAs (CHILD\_SA\_REKEY). Under the Make Before Break (MBB) discipline, the new SA is installed before the old one is deleted, so a rekey only endangers the tunnel if the handshake outlives the hard lifetime window.

\subsection{Multiple KEs in IKEv2}
Two IETF extensions make it possible to establish hybrid keys. RFC 9242~\cite{rfc9242} introduces the IKE\_INTERMEDIATE exchange, which can carry large payloads after IKE\_SA\_INIT but before IKE\_AUTH, using IKE level fragmentation. RFC 9370~\cite{rfc9370} builds on it and allows up to seven additional KEs. During rekeying the additional exchanges travel in IKE\_FOLLOWUP\_KE messages. The session key mixes all negotiated secrets, so the tunnel remains secure as long as at least one method resists~\cite{rfc9370}. This is the mechanism that lets us combine X25519, ML-KEM, and a QKD derived key in one handshake. A complementary approach, which mixes PQC preshared keys into the exchanges, was recently standardized in RFC 9867~\cite{rfc9867}.

\subsection{ML-KEM and QKD Key Delivery}
NIST standardized ML-KEM (FIPS 203~\cite{fips203}) as the primary PQC key encapsulation mechanism that consists of only a software layer. However, QKD systems produce symmetric keys at both ends of a quantum link and applications access them through the local KME via the ETSI GS QKD 014 REST interface~\cite{etsi014}, where the initiator obtains a key and identifier that the responder uses to retrieve the matching key.

\section{Related Work}
\label{sec:sota}

The closest work to ours is by Blanco-Romero \emph{et al.}~\cite{blanco2025hybrid}, who presented a systematic comparison of sequential (RFC 9370 style) and parallel hybrid QKD and PQC key establishment for IKEv2, supporting both the ETSI 004 and ETSI 014 APIs. Their central result is that every additional sequential exchange costs one full network round trip, which becomes a multiplicative penalty on high latency paths under an artificial delay of 100~ms. Their sequential QKD hybrid needed about 200~ms more than the parallel variant. Our results complement theirs from the other side of the trade off: on a metropolitan link the extra QKD exchange costs only about 7~ms, so the sequential composition mandated by RFC 9370 is affordable exactly where QKD hardware is deployed today, namely metropolitan distances. More important, neither their work nor any other in Table~\ref{tab:related} addresses what happens when the QKD key delivery fails during a handshake. To the best of our knowledge, this paper is the first to design, implement, and experimentally validate graceful degradation for a hybrid PQC and QKD IKEv2 tunnel.

\section{System Design}
\label{sec:design}

\begin{figure*}[!t]
    \centering
    \includegraphics[width=\textwidth]{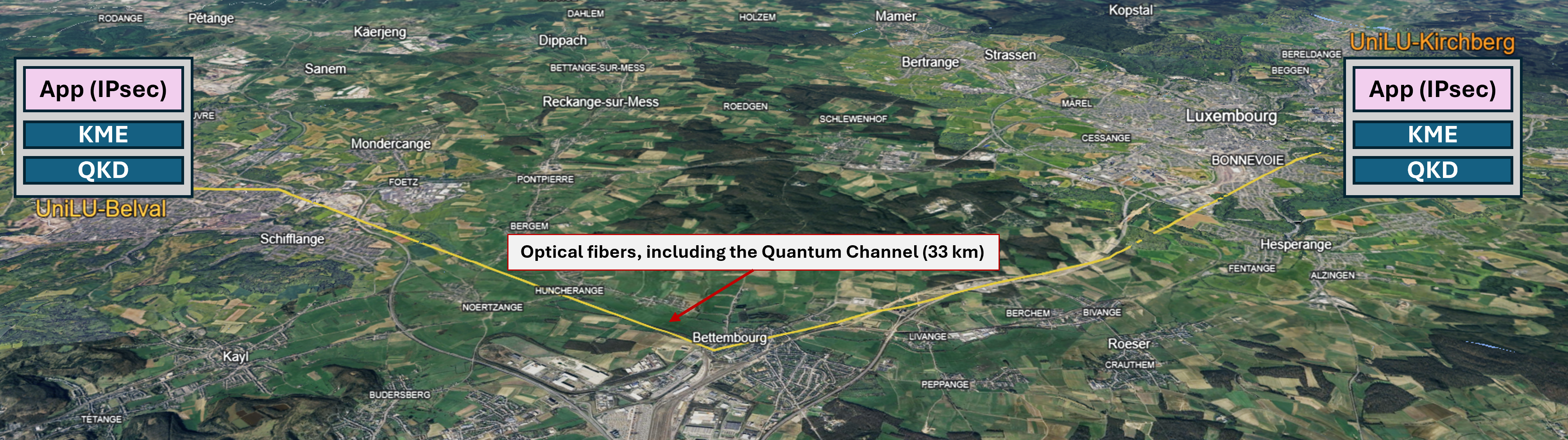}
    \caption{Testbed used in the experiments. Two sites host an IPsec endpoint, a KME, and a QKD device each. The QKD systems are connected through deployed optical fiber that includes a 33 km quantum channel within LUQCIA.}
    \label{fig:testbed}
\end{figure*}

\subsection{Architecture Overview}
Our system consists of two IPsec peers, an initiator and a responder, that protect traffic between two sites within the LUQCIA (LUxembourg Quantum Communication Infrastructure) infrastructure, as depicted in Fig.~\ref{fig:testbed}. Each site hosts three elements: the IPsec endpoint itself, a local KME, and a QKD device from IDQ. The two QKD devices are connected by a quantum channel and produce identical symmetric keys at both sites. The KMEs store these keys and serve them to applications through standard interfaces~\cite{etsi014}. The IPsec endpoints run strongSwan extended with our QKD plugin and together with the OpenSSL provider that supplies the ML-KEM implementation. The IKEv2 negotiation uses three key exchange components: X25519 as the classical component in IKE\_SA\_INIT, ML-KEM as the first additional key exchange (KE), and the QKD derived key as the second additional KE, both negotiated through RFC 9370. All three secrets enter the IKEv2 key derivation.

\subsection{The QKD Plugin}
The plugin implements the QKD component for strongSwan, so no change is needed to the IKEv2 state machine. During the QKD KE round, the initiator calls the \texttt{Get key} method of its local KME and obtains a 256 bit key together with its identifier (a UUID). Only the identifier travels inside the KE payload. The key itself never crosses the classical network. The responder presents the received identifier to its own KME with the \texttt{Get key with key IDs} method and obtains the same key. Both peers then feed the key into the standard RFC 9370 key derivation. 

\subsection{Graceful Degradation}
\label{sec:degradation}

\begin{figure*}[!t]
    \centering
    \includegraphics[width=0.65\textwidth]{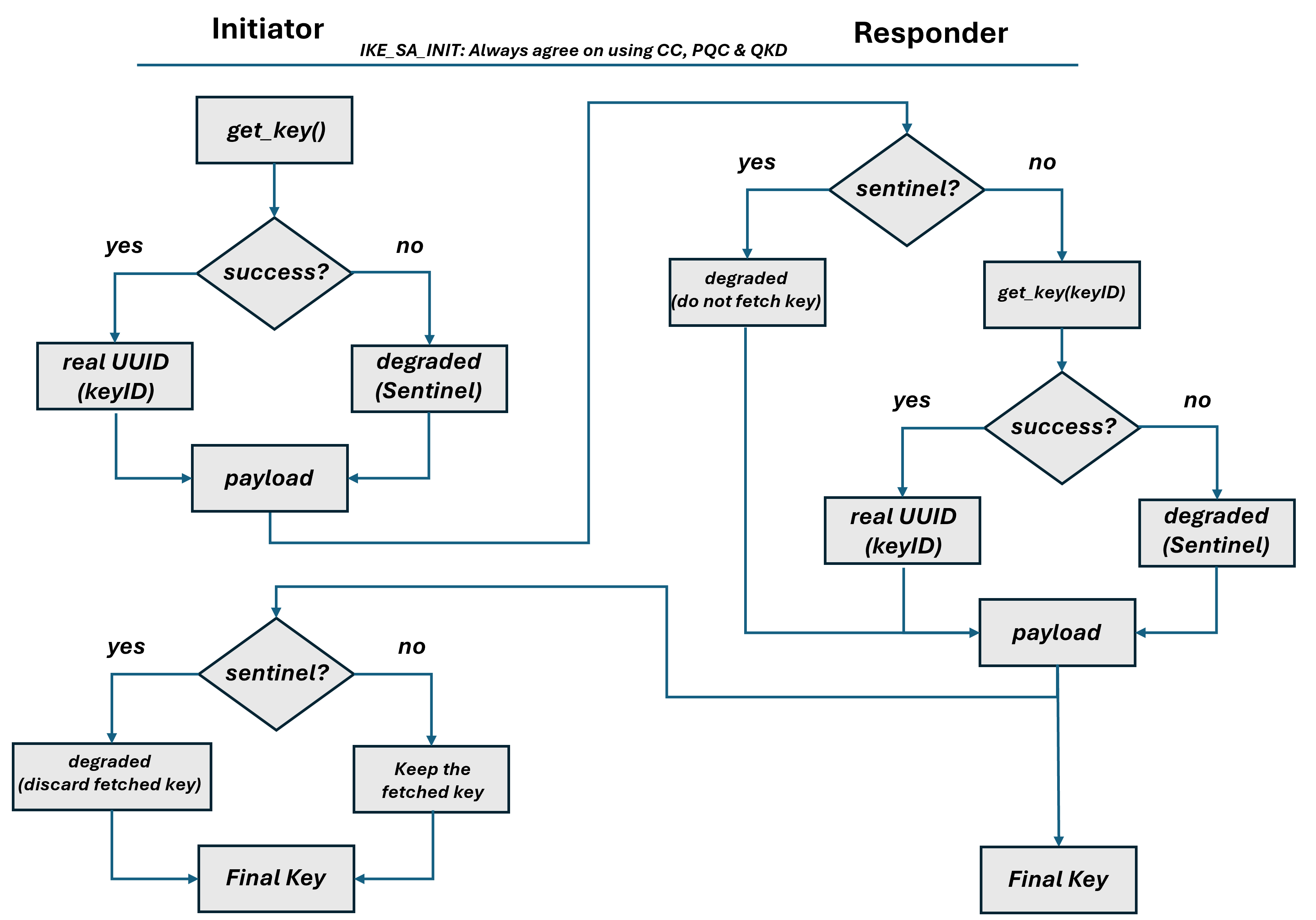}
    \caption{Sentinel-based graceful degradation protocol. Both peers evaluate the availability of their KME inside the normal QKD KE round and converge on the same decision: use the QKD share, or complete the handshake without it.}
    \label{fig:degradation}
\end{figure*}

Assuming that the QKD branch can fail, a naive integration would also make the IKE\_FOLLOWUP\_KE exchange fail and, with it, the whole negotiation. Our design goal is the following: the tunnel must always be established, with the strongest set of components that is available at that moment, and it must upgrade itself again as soon as the QKD service returns in the next KE.

Fig.~\ref{fig:degradation} shows the protocol. During IKE\_SA\_INIT the peers always agree on the full proposal, classical plus PQC plus QKD, in such way that the degradation decision is taken per exchange and not per SA. In the QKD round, the initiator calls \texttt{get\_key()} on its KME. On success it places the real key identifier in the KE payload. In case of failure it places a sentinel value, an all zero identifier that can never collide with a real UUID. The responder inspects the payload first. If it finds the sentinel, it does not contact its own KME at all and answers with the sentinel as well. The QKD share is skipped on both sides. If it finds a real identifier, it calls \texttt{get\_key(keyID)}. On success it confirms with the real identifier, and on failure it answers with the sentinel. Finally, the initiator inspects the response. If the response carries the sentinel, it discards the key it had fetched and both peers derive the final key without the QKD share. Otherwise both keep the fetched key. In every branch the two peers reach the same decision using only the payloads of the exchange itself, so no extra round trip, no error notification, and no renegotiation are needed. We have set a maximum time of 2 seconds to wait for the KME responses. Otherwise, the KME is assumed out of service.  

Security wise, a degraded handshake still combines X25519 and ML-KEM, so it never falls below the post-quantum security floor. Because the decision is repeated at every rekey, the recovery is automatic: at the first IKE\_SA\_REKEY after the KME returns, the \texttt{get\_key()} call succeeds again. The maximum time in degraded mode is therefore bounded by the rekey interval, which the operator can tune.

\section{Experimental Setup}
\label{sec:setup}

\subsection{Evaluated Configurations}
We evaluate five IKEv2 configurations, labeled A to E, that move step by step from a fully classical handshake to a hybrid quantum-safe one as defined in Table~\ref{tab:configs}. All five configurations authenticate with RSA certificates, so any difference between them comes only from the key establishment. ESP traffic is protected with AES-256-GCM in all cases. We have assumed that only the initiator can trigger the rekey.

\begin{table}[!t]
\caption{IKEv2 Configurations Evaluated}
\label{tab:configs}
\centering
\begin{tabular}{cll}
\toprule
\textbf{Scenario} & \textbf{Key Exchange(s)} & \textbf{Authentication} \\
\midrule
A & X25519 & RSA \\
B & X25519 + ML-KEM-768 & RSA \\
C & X25519 + ML-KEM-1024 & RSA \\
D & X25519 + ML-KEM-768 + QKD & RSA \\
E & X25519 + ML-KEM-1024 + QKD & RSA \\
\bottomrule
\end{tabular}
\end{table}

\subsection{Measurement Methodology}
Every IKEv2 exchange is measured as a request to response round trip, taken from the strongSwan event timestamps on the initiator. For each configuration we run repeated cycles of tunnel establishment, IKE\_SA\_REKEY, and CHILD\_SA\_REKEY, and we separate the first handshake of each run (cold start, when plugins, random number generators, and certificate caches are still empty) from the steady state ones. In addition, we capture the traffic between the peers with tcpdump to verify the protocol behavior on the wire, and we collect the charon daemon logs on both sides for the failure injection experiments. The KME outage is injected by making the KME service unreachable and/or key exhausted on any side of the tunnel.

\section{Experimental Results}
\label{sec:results}

\subsection{Protocol Behavior on the Wire}

\begin{figure*}[!t]
    \centering
    \includegraphics[width=.75\textwidth]{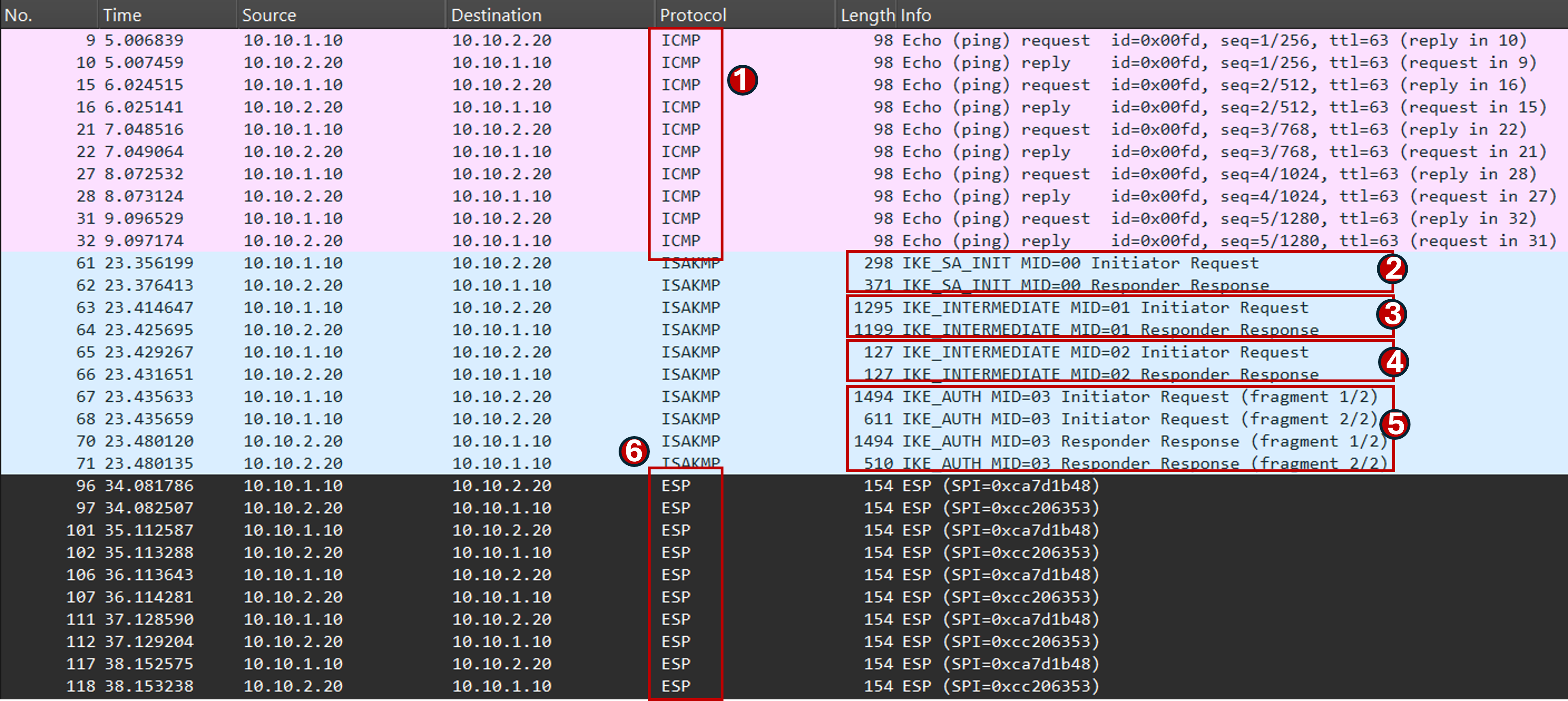}
    \caption{Packet capture of a hybrid handshake. ICMP traffic was used (5 pings before the tunnel and 5 more after the tunnel was up.)}
    \label{fig:pcap}
\end{figure*}

Fig.~\ref{fig:pcap} shows a capture of a complete handshake in configuration D according to Table~\ref{tab:configs}. The IKE\_SA\_INIT pair stays small (298 and 371 bytes) because it carries only the X25519 public values. The first IKE\_INTERMEDIATE exchange transports the ML-KEM-768 material and grows to 1295 and 1199 bytes, close to the Ethernet MTU. The second IKE\_INTERMEDIATE exchange is the QKD round: both datagrams are only 127 bytes, because the peers exchange a key identifier and not the key itself. The IKE\_AUTH exchange, which carries the certificates, is the only one that needs IKE level fragmentation, with two datagrams per direction. After the last response, the ESP packets show up and the traffic between the two hosts is protected.

\subsection{Handshake and Rekey Cost}

\begin{figure*}[!t]
    \centering
    \includegraphics[width=0.9\textwidth]{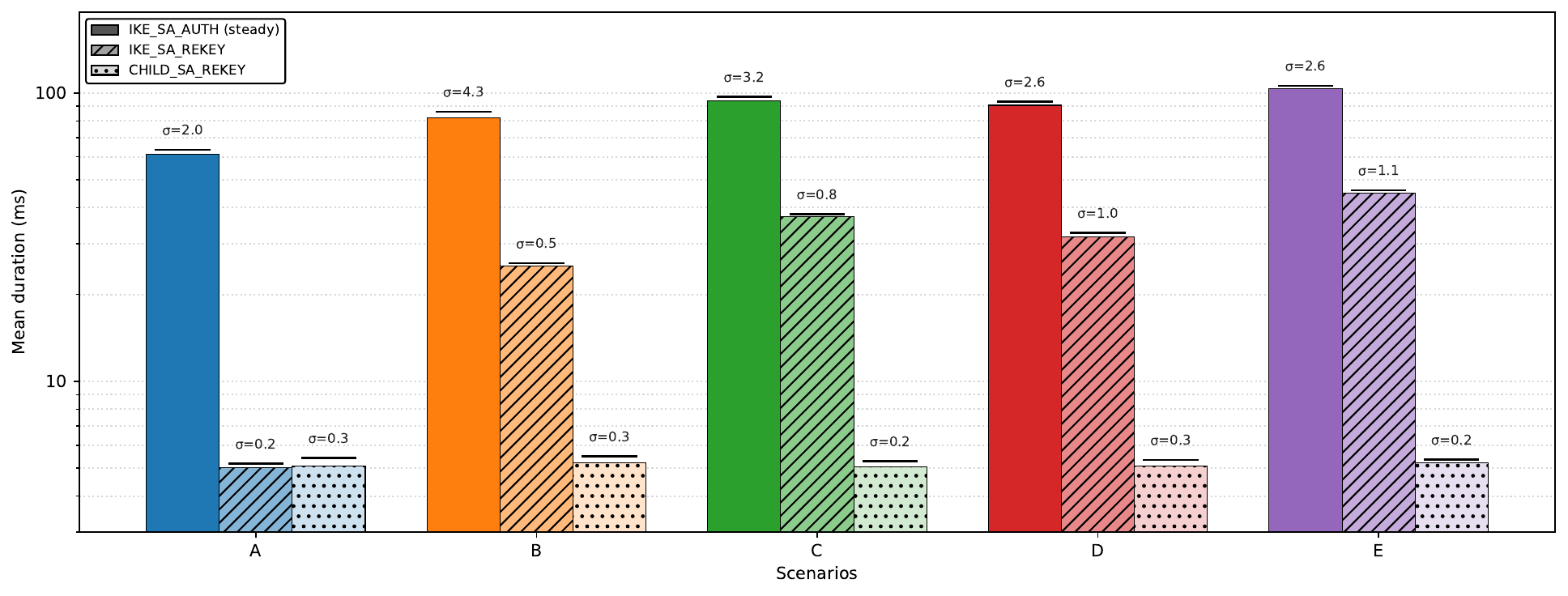}
    \caption{Mean duration of the IKE\_SA\_AUTH (steady state). The annotations are the standard deviation of each bar.}
    \label{fig:cost}
\end{figure*}

Fig.~\ref{fig:cost} reports the mean duration of the three exchange types for the five configurations. The complete authentication grows from 61 ms in the classical baseline A to 103 ms in the full hybrid E, an increase of a factor 1.7. The dispersion is small in every case as the standard deviation stays between 2.0 and 4.3 ms for the authentication and below 1.1 ms for the rekeys, so no heavy tail hides behind the means.

The IKE\_SA\_REKEY row isolates the pure key establishment cost, because a rekey repeats the key exchange without certificates. Adding ML-KEM-768 lifts the rekey from 5 ms to 25 ms, and ML-KEM-1024 pushes it to 37 ms, partly because its public key no longer fits in one UDP datagram and the exchange fragments. Adding the QKD round on top costs about 7 ms more. This 7 ms covers one extra round trip on our link plus the two REST calls to the KMEs, so the ETSI 014 key fetch itself is nearly free. This observation also quantifies, for a metropolitan deployment, the sequential composition penalty analyzed by Blanco-Romero \emph{et al.}~\cite{blanco2025hybrid}: the extra exchange costs exactly one round trip, which is negligible here and grows only with the network latency.

Finally, CHILD\_SA\_REKEY stays at about 5 ms in all five configurations because it only involves key derivation from previously established key sessions.

\subsection{Where the Milliseconds Go}

\begin{figure}[!t]
    \centering
    \includegraphics[width=\columnwidth]{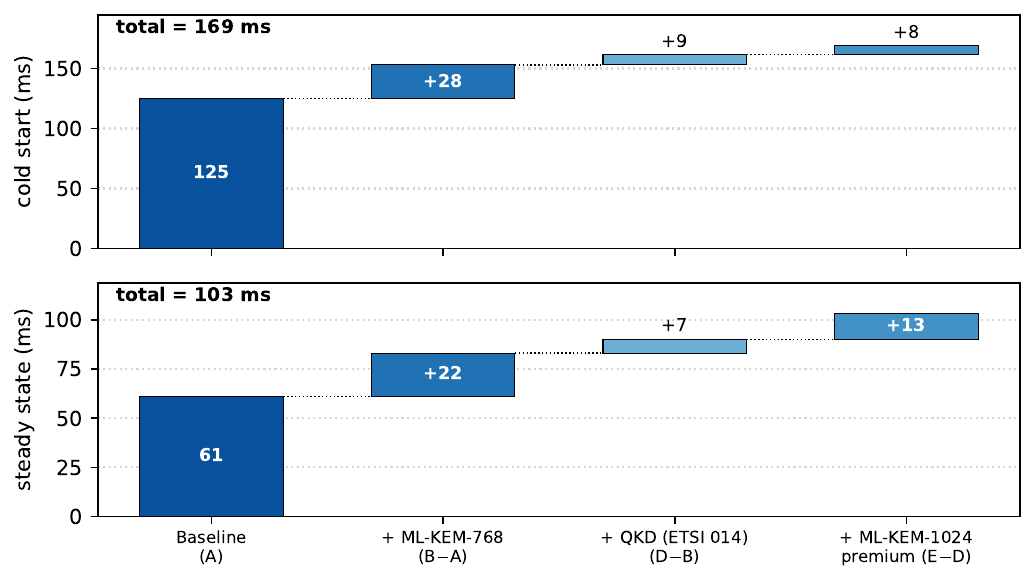}
    \caption{Incremental cost of the hybrid components on the complete authentication, at cold start (top) and in steady state (bottom).}
    \label{fig:waterfall}
\end{figure}

Fig.~\ref{fig:waterfall} decomposes the different configurations into incremental steps, both in the cold start and in steady state. In steady state, the baseline costs 61 ms, ML-KEM-768 adds 22 ms, the QKD round adds 7 ms. Moving from ML-KEM-768 to ML-KEM-1024 adds a final 13 ms, for a total of 103 ms. At cold start the picture is similar but shifted upward: the baseline alone costs 125 ms, and the full hybrid reaches 169 ms. The cold start surplus comes from one time effects, mainly plugin and random number generator initialization and the first validation of the certificate chain, so it is a commissioning event rather than a steady state concern. One conclusion is clear: the QKD component is the cheapest of the three additions, which removes the performance argument against including it wherever a QKD link exists.

\subsection{Failure Injection and Recovery}

\begin{figure*}[!t]
    \centering
    \includegraphics[width=0.85\textwidth]{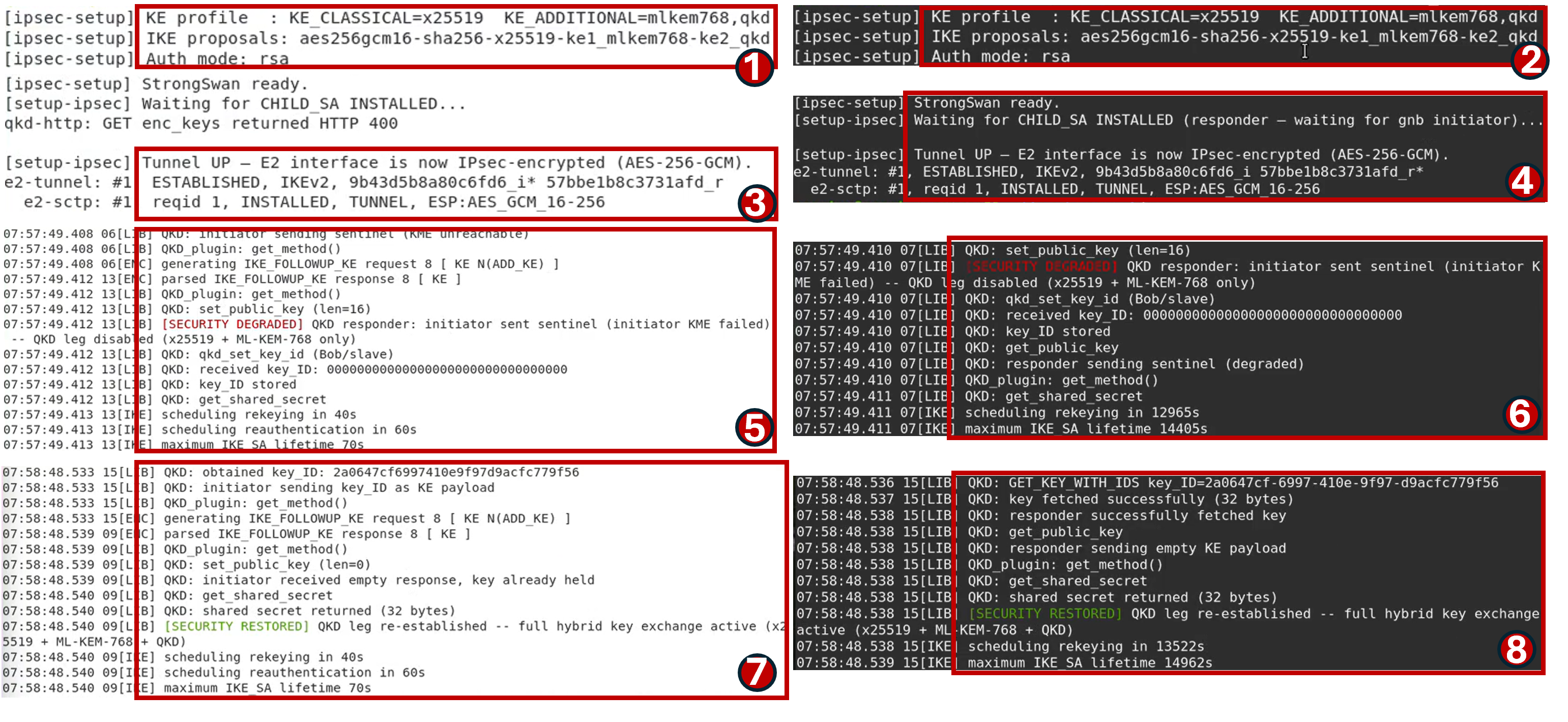}
    \caption{Daemon logs of a full degradation and recovery cycle, initiator on the left and responder on the right.}
    \label{fig:logs}
\end{figure*}

Fig.~\ref{fig:logs} shows a complete degradation and recovery cycle as observed in the charon logs of both peers. The tunnel is first established with the full hybrid profile (see points 1 to 4). We then insert a failure in the KME on the initiator side and wait for the next IKE\_SA\_REKEY. The initiator \texttt{get\_key()} call fails, so it sends the sentinel identifier ( 0x0000). The responder recognizes it, skips its own KME call, and both daemons log the event with the \texttt{SECURITY DEGRADED} label, explicitly saying that the QKD leg is disabled and that the exchange continues with X25519 and ML-KEM-768 only (points 5 and 6). The rekey completes normally and the ESP traffic is never interrupted. From the point of view of the protected applications, nothing happens. After we restore the KME, the following rekey fetches a fresh 32 byte key through \texttt{Get key with key IDs}, both daemons log \texttt{SECURITY RESTORED}, and the full hybrid key exchange is active again with the next rekey already scheduled (points 7 and 8), making evident that the initiator is the one that fires the rekey thanks to having the lowest values for the SA lifetime.  

Three properties are worth noting. First, the degradation is decided inside the normal message flow, so it adds no extra exchange and no failure specific latency. Second, the mechanism is symmetric: the same sentinel logic covers an outage on the responder side, where the responder answers with the sentinel and the initiator discards its fetched key, keeping the two key derivations consistent. Third, the exposure window is bounded by the rekey interval, so the operator controls directly how long the tunnel may stay without its QKD share.

\section{Conclusion}
\label{sec:conclusion}
We presented the design, implementation, and evaluation of a quantum-safe key establishment scheme for IPsec tunnels that combines X25519, ML-KEM, and QKD keys in a single IKEv2 negotiation and that degrades gracefully when the QKD leg fails. On a testbed attached to a metropolitan QKD link over 33 km of deployed fiber, the full hybrid key establishment raises the complete authentication from 61 ms to 103 ms, the QKD round itself costs only about 7 ms thanks to its identifier based design, and the frequent child rekeys stay at about 5 ms in every configuration. The sentinel based degradation protocol keeps the tunnel alive through a complete KME outage without interrupting the protected traffic, never falls below the post-quantum security floor, and restores the QKD share automatically at the next rekey. Together, these results remove the two main practical objections against QKD in IPsec: the performance cost is small, and the availability risk can be engineered away. Future works will focus on extending the evaluation to post-quantum authentication methods and preshared key modes such as RFC 9867~\cite{rfc9867}. We also plan to apply and evaluate the tunnel in specific critical infrastructures.

\section*{Acknowledgment}
This work was supported by XTRUST-6G and LUQCIA projects. XTRUST-6G is co-funded by the European Union. Views and opinions expressed are however those of the author(s) only and do not necessarily reflect those of the European Union or Smart Networks and Services Joint Undertaking. Neither the European Union nor the granting authority can be held responsible for them. This work has received funding from the Swiss State Secretariat for Education, Research and Innovation (SERI). LUQCIA is funded by the European Union - Next Génération EU, with the collaboration of the Department of Média, Connectivity and Digital Policy in the framework of the project LUQCIA.

\bibliographystyle{IEEEtran}
\bibliography{ref}

@article{blanco2025hybrid,
  author  = {Blanco-Romero, J. and Otero Garcia, P. and Sobral-Blanco, D. and Almenares Mendoza, F. and Fernandez Vilas, A. and Fernandez-Veiga, M.},
  title   = {Hybrid Quantum Security for {IPsec}},
  journal = {arXiv:2507.09288},
  year    = {2025}
}

@inproceedings{twardokus2025viability,
  author    = {Twardokus, Geoff and Joslin, William and Rahbari, Hanif and Layton, William},
  title     = {Assessing the Viability of Quantum-Resistant {IKEv2} over Constrained and Internet-Scale Networks},
  booktitle = {Proc. 2025 Quantum Security and Privacy Workshop (QSec '25)},
  address   = {Taipei, Taiwan},
  publisher = {ACM},
  year      = {2025},
  doi       = {10.1145/3733825.3765281}
}

@inproceedings{herzinger2023realworld,
  author    = {Herzinger, Daniel and Gazdag, Stefan-Lukas and Loebenberger, Daniel},
  title     = {Real-World Quantum-Resistant {IPsec}},
  booktitle = {Proc. 14th International Conference on Security of Information and Networks (SIN)},
  publisher = {IEEE},
  year      = {2021}
}

@article{iliadis2025qrons,
  author  = {Iliadis-Apostolidis, Dimosthenis and Lawo, Daniel Christian and Kosta, Sokol and Tafur Monroy, Idelfonso and Vegas Olmos, Juan Jose},
  title   = {{QRoNS}: Quantum Resilience over {IPsec} Tunnels for Network Slicing},
  journal = {Electronics},
  volume  = {14},
  number  = {21},
  pages   = {4234},
  year    = {2025},
  doi     = {10.3390/electronics14214234}
}

@article{vicente2026banking,
  author  = {Vicente, Rafael J. and G{\'o}mez Garc{\'i}a, Jaime and Brito, Juan P. and Lobaina, Yorlandy and Buruaga, Jaime S. and G{\'o}mez Aguado, Daniel and S{\'a}nchez Serrano, Miguel {\'A}ngel and Ovsyannikov, Sim{\'o}n and Gherdaoui, Salah and Pegon, Jean-S{\'e}bastien and Cofano, Marco and Mart{\'i}n, Vicente},
  title   = {Quantum-Safe {IPsec} in the Banking Industry},
  journal = {arXiv preprint arXiv:2604.12985},
  year    = {2026}
}

@article{alia2024deployedfiber,
  author  = {Alia, Obada and Huang, Albert and Luo, Huan and Amer, Omar and others},
  title   = {100 {Gbps} Quantum-Safe {IPsec} {VPN} Tunnels over 46 km Deployed Fiber},
  journal = {arXiv preprint arXiv:2405.04415},
  year    = {2024}
}

@article{dervisevic2021qkdipsec,
  author  = {Dervisevic, E. and Mehic, M.},
  title   = {Overview of Quantum Key Distribution Technique within {IPsec} Architecture},
  journal = {arXiv:2112.13105},
  year    = {2021}
}

@misc{rfc7296,
  author       = {Kaufman, Charlie and Hoffman, Paul and Nir, Yoav and Eronen, Pasi and Kivinen, Tero},
  title        = {Internet Key Exchange Protocol Version 2 ({IKEv2})},
  howpublished = {IETF RFC 7296},
  year         = {2014},
  doi          = {10.17487/RFC7296}
}

@misc{rfc9242,
  author       = {Smyslov, Valery},
  title        = {Intermediate Exchange in the Internet Key Exchange Protocol Version 2 ({IKEv2})},
  howpublished = {IETF RFC 9242},
  year         = {2022},
  doi          = {10.17487/RFC9242}
}

@misc{rfc9370,
  author       = {Tjhai, C. J. and Tomlinson, M. and Bartlett, G. and Fluhrer, S. and Van Geest, D. and G-Morchon, O. and Smyslov, V.},
  title        = {Multiple Key Exchanges in the Internet Key Exchange Protocol Version 2 ({IKEv2})},
  howpublished = {IETF RFC 9370},
  year         = {2023},
  doi          = {10.17487/RFC9370}
}

@misc{rfc9867,
  author       = {Smyslov, Valery},
  title        = {Mixing Preshared Keys in the {IKE\_INTERMEDIATE} and {CREATE\_CHILD\_SA} Exchanges of {IKEv2} for Post-Quantum Security},
  howpublished = {IETF RFC 9867},
  year         = {2025},
  doi          = {10.17487/RFC9867}
}

@misc{fips203,
  author       = {{National Institute of Standards and Technology}},
  title        = {Module-Lattice-Based Key-Encapsulation Mechanism Standard},
  howpublished = {NIST FIPS 203},
  year         = {2024},
  doi          = {10.6028/NIST.FIPS.203}
}

@misc{etsi014,
  author       = {{ETSI}},
  title        = {Quantum Key Distribution ({QKD}); Protocol and Data Format of {REST}-Based Key Delivery {API}},
  howpublished = {ETSI GS QKD 014 V1.1.1},
  year         = {2019}
}

\end{document}